\documentclass{elsart4-1}

\usepackage{amssymb}

\usepackage[final]{LLNdef}
\usepackage{journames}
\usepackage[final]{modifications}
\usepackage{cite}
\usepackage{citesort}

\usepackage[english,francais]{babel}

\newtheorem{e-proposition}[theorem]{Proposition}

\newtheorem{e-definition}[theorem]{Definition\rm}

\def\og{\leavevmode\raise.3ex\hbox{$\scriptscriptstyle\langle\!\langle$~}}
\def\fg{\leavevmode\raise.3ex\hbox{~$\!\scriptscriptstyle\,\rangle\!\rangle$}}

\begin{document}

\begin{frontmatter}


\selectlanguage{english}
\title{Self-organization on surfaces: foreword}

\vspace{-2.6cm}

\selectlanguage{francais}
\title{Auto-organisation sur les surfaces~: pr{\'e}face}


\selectlanguage{english}
\author[of]{Olivier Fruchart}
\ead{Olivier.Fruchart@grenoble.cnrs.fr}

\address[of]{Laboratoire Louis N{\'e}el -- 25, avenue des Martyrs -- BP166 -- F-38042 Grenoble Cedex 9 -- France}

\begin{abstract}
After decades of work, the growth of continuous thin films, \ie, two-dimensional structures, is
progressively becoming a technological issue more than a field of fundamental research.
Incidentally self-organization of nanostructures on surfaces is now an important field of
research, \ie, structures of dimensionality one or zero, with a steep rise of attention in the
past five years. Whereas self-organization was initially motivated by potential applications, it
has up to now essentially contributed to the advancement of fundamental science in low dimensions,
as model systems could be produced that could not have been fabricated by lithography. This
Special Issue aims at giving a cross-community timely overview of the field. The Issue gathers a
broad panel of articles covering various self-organization mechanisms, specific structural
characterization, physical properties, and current trends in extending the versatility of growth.
The materials mostly covered here are semiconductors and magnetic materials. {\it To cite this
article: O. Fruchart, C. R. Physique 6 (2005).}

\vskip 0.5\baselineskip

\selectlanguage{francais}
\noindent{\bf R{\'e}sum{\'e}}%
\vskip 0.5\baselineskip%
\noindent%
Apr{\`e}s plusieurs d{\'e}cennies d'intense activit{\'e}, la croissance des couches minces, \cad de syst{\`e}mes
de dimension deux, tend progressivement {\`a} devenir plus une question de ma{\^\i}trise technologique
qu'un sujet de recherche fondamentale. Dans le m{\^e}me temps l'auto-organisation aux surfaces de
nanostructures, \cad de dimensionnalit{\'e} un voire z{\'e}ro, prend une importance grandissante, tout
sp{\'e}cialement depuis cinq ans. Alors que l'auto-organisation {\'e}tait initialement motiv{\'e}e par des
perspectives d'applications, elle a pour l'instant essentiellement contribu{\'e} {\`a} l'acquisition de
nouvelles connaissances de physique fondamentale en basse dimension, puisque des syst{\`e}mes mod{\`e}les
ont pu {\^e}tre fabriqu{\'e}s, qui n'auraient pu {\^e}tre obtenus par la technique conventionnelle de
structuration qu'est la lithographie. Ce dossier th{\'e}matique a pour objectif de proposer un aper\c{c}u
actuel de l'auto-organisation, en essayant de d{\'e}passer les fronti{\`e}res de communaut{\'e}s. Ainsi sont
rassembl{\'e}es des contributions traitant de divers m{\'e}canismes physiques de l'auto-organisation, de
techniques de caract{\'e}risation sp{\'e}cifiques, de propri{\'e}t{\'e}s physiques, et des nouvelles approches
poursuivies pour augmenter la versatilit{\'e} des proc{\'e}d{\'e}s de croissance. Les mat{\'e}riaux couverts ici
sont essentiellement les semiconducteurs et les mat{\'e}riaux magn{\'e}tiques. {\it Pour citer cet
article~: O. Fruchart, C. R. Physique 6 (2005).}

\vskip 0.5\baselineskip%
\noindent{\small{\it Mots-cl{\'e}s~:} Auto-organisation~; Auto-assemblage~; Bottom-up; Basse dimensionnalit{\'e}, Plots~; Fils~; Nanostructures}%
}\end{abstract}
\end{frontmatter}

\selectlanguage{english}

\section{Introduction}
\label{}

A draft definition of self-organization~(\SO) could be: \emph{a situation were structures with a
certain degree of order appear spontaneously}, or for the restricted case of growth at a surface:
\emph{a growth process during which the spontaneous formation of nanostructures occurs}. For
decades growth parameters and the combination of materials have been optimized to yield continuous
thin films of the highest structural quality and smoothness upon growth at surfaces; lithography
was then used when circuits and other patterns of finite lateral dimensions were required. On the
contrary, materials and growth parameters can be optimized with an opposite goal, that of
producing spontaneously nanostructures on a surface. This is self-organization, which appears as
an alternative route to lithography. Self-organization can be viewed as the generalization of
growth processes to dimensions lower than two. Indeed thin films growth makes use of the
aggregation of material at a surface~(2D), while self-organization makes use of aggregation along
lines~(1D) and points~(0D) to yield stripes and dots. It is the purpose of this Special Issue to
examine the current status of self-organization at surfaces and its uses.

The first motivation put forward to study self-organization is often the potential for hi-tech
applications. Indeed, the exponential increase of integration and performance of devices is
accompanied by a similar increase of the cost of investment for building the fabrication
facilities. Self-organization, that allows the spontaneous fabrication of nanostructures upon a
single and cheap growth procedure has attracted attention for its potential to become an
alternative route to lithography to push further the limits of integration at a reasonable cost.
The increase of density of lithographically-defined patterns is called a \emph{top-down} approach,
were scientists and engineers work on improving the concepts in use for today-lithography.
Alternative approaches are called \emph{bottom-up}, were one tries to tailor the growth of
individual entities, like atoms, clusters, molecules\ldots, up the to spontaneous fabrication of
patterns. However, although some serious prospects still exists, there has been no breakthrough
for applications after nearly two decades of research. Intense work is nonetheless still going on
in the field, pushed by the potential for fabricating model systems for fundamental studies, that
could not be achieved by any other existing technique. We will see that the effort in the field
has lead to the advancement of physics in low dimensions. Notice that technology and fundamental
research in the nano-world are nonetheless closely related, as new effects are expected at the
scale of nanometers. As such effects cannot be predicted from the properties of larger objects, it
would not be desirable to have investigations of self-organization guided solely by the need to
fabricate in the low-nanometer scale patterns whose functionality are desired and known at a
larger scale. Fundamental research is a mandatory step to explore the field in all directions with
no \apriori prospect, and is a sufficient justification for the current effort in the field of
self-organization.

Why is today an overview of \SO timely? \SO first came in the mid-eighties from the semiconductor
community, with  the technological prospect of high-efficiency lasing in quantum
dots\cite{bib-ARA82,bib-ASA86}. The rapid expansion of the field was facilitated by the incidental
introduction at the same time of scanning probe microscopies like scanning tunneling microscopy
and atomic force microscopy. Now \SO is a broad and established field of epitaxial growth in many
communities, both in terms of materials and physics. In the past five years many new ramifications
emerged beyond the study of growth and of specific properties like luminescence for
semiconductors, magnetic anisotropy for magnetic materials, \etc. For example X-ray
characterization techniques specific to \SO were developed. At the same time the growth
engineering was still being pushed further. Investigated directions include a bottom-up approach
to fabricate truly 3D materials now that the structure at the nano-level is controlled at
surfaces, or combine a first versatile step of pre-patterning, \eg, lithography, with overgrowth.
It is therefore timely to propose an overview of what has been achieved, and of current prospects
of technological and fundamental interest. The fact that \SO is now spread in many communities has
motivated a broad overview of the field in this Special Issue, rather than a set of very
specialized reviews on a narrow subject. Although some specialized reviews about \SO were
available\cite{bib-BRU98b,bib-HEN98,bib-TEI02,bib-ROU02b,bib-SHC99,bib-TRO02}, mainly devoted to
growth, a broad set of reviews was not available to date. Indeed, as self-organization requires
the engineering of growth processes, and also accurate characterization techniques, there is often
a lot to gain at looking at other communities than our own's. It is also essential to understand
what kind of new physical functionalities might arise from \SO systems to tailor the growth in the
good direction. With these ideas, I hope that this Issue will help the reader to gain an overview
of the field of \SO.

Let us first discuss a semantic issue. We will see that two prototypical situations are
encountered concerning the spontaneous formation of nanostructures upon growth: the nanostructures
may either display only weak nearest-neighbor correlations, or display a mid- or long-range
positional order. It seem therefore desirable to use different names to distinguish these two
situations. Incidentally two names are found in the literature to describe the spontaneous
formation of nanostructures on surfaces, \ie, \emph{self-organization}~(\SO) and
\emph{self-assembly}~(\SA)\cite{bib-CRphys-editor-SOSA}. However there is no agreement in the
community as to the exact use of these two terms, that are sometimes used interchangeably, or
whose meaning may depend on the community. To reconcile this lack of consensus with the need for
two distinct names stressed above, I propose to call \emph{self-organization} the prototypical
case of nanostructures displaying a mid- or long-range positional order, and \emph{self-assembly}
the prototypical case of nanostructures displaying only weak nearest-neighbor positional
correlations. Of course, in real situations there may be no strict borderline between these two
situations. If one forgets about this issue of positional order, the use of \SO and \SA
nanostructures can be classified into two categories. In the first category they are used to gain
information about fundamental phenomena that occur in materials and systems that may be of
interest for applications, but cannot be understood directly, because they are too complex~(owing
to microstructure, defects, large size \ldots). Beyond the advancement of our understanding of
physics in low-dimension, these studies are of applied interest, because devices require the use
of ever smaller nanostructures, whose properties must be understood and ultimately tailored. In
this case \SO systems are used preferably to lithography or microstructured materials as objects
of very high quality to serve as model systems for analyzing fundamental issues of physics.
Indeed, these are characterized by a very high resolution, ultimately down to the atomic
size\cite{bib-GAM02b,bib-GAM03}, much better than existing and presumably any forthcoming
lithography technology. Besides, the quality of the interfaces can be potentially controlled at
this scale, with presumably much less defects than for lithography~(roughness, amorphisation,
oxydation or other gas adsorbtion \ldots). Thus, they possess the required features to study
fundamental physical phenomena in low dimensions, as the use of a model system maximizes the
chances to elucidate the relevant phenomena quantitatively. The second category of studies
consists in investigating whether \SO systems might be used directly for applications. This still
remains an open question after twenty years of research: the physical properties of these systems
are often appealing, but technological barriers remain to be overcome. These two categories of
studies will be illustrated in this Special Issue. Finally, notice that \SO receives often more
attention than \SA. This partly stems from the fascination for an order that can arise
spontaneously. However from scientific and technological points of view the advantage of \SO\ most
often is not the order itself, but the fact that in ordered patterns the size distribution is very
narrow, because the area of capture of all objects during growth is similar. For fundamental
science this gives researchers the opportunity to assess the properties of nano-sized systems by
measuring large assemblies of these, and assuming that all entities contribute in a similar way.
From a technological point of view, reducing the size dispersion is important to reduce the
dispersion of physical properties of use in a device, like wavelength in a quantum-dot laser or
coercive field in a magnetic recording media. The sketch of the Issue is given in the following.

\vskip 0.5in

The first article, by Olivier \surname{Pierre-Louis}, is a review of the literature work about
\emph{steps on surfaces}\cite{bib-CRphys-Pierre-Louis}. As crystals can never be polished exactly
along surfaces of low Miller indices, steps are always found on surfaces to compensate for the
miscut angle. Thus the study of steps on surfaces is a long-standing issue in surface
physics\cite{bib-BUR51}, and steps have been observed to self-assemble or self-organize.
Experiments were first made possible by electron microscopies like grazing incidence electron
microscopy, or by LEEM, and then revived by the introduction of scanning probe microscopies.
Experimentally parameters like the miscut angle and growth conditions provide theoreticians with
sets of data to help understanding both thermodynamic and kinetic processes. Despite the large
number of models and simulations released especially in the last fifteen years, the subject is
still timely. The current trend is the increasing overlap of phenomenological models with
microscopic aspects like strain and electronic structure in a multi-scale approach. The knowledge
of the base models in this well-established field is worth, as the microscopic ingredients
governing the interaction between steps are also often those who are responsible for the \SO and
\SA of nanostructures deposited on surfaces.

Another many-decades-old issue on crystalline surfaces is the classification of growth
modes\cite{bib-VOL26,bib-STR38,bib-FRA49}, that was first rationalized by Bauer in
1958\cite{bib-BAU58} in the growth modes now called Volmer-Weber, Stranski-Krastanov, and
Frank-van-den-Merwe. For reviews, see for instance \cite{bib-ZAN88,bib-BRU01,bib-PIM99}. However
Nature does not like drawers and boxes, and models need always to be refined to describe different
systems, physical effects, and dimensions of the patterns. Thus there is still a significant
activity going on to explain and then tailor the growth of nanostructures, especially concerning
the conditions of occurrence and the shape of dots\cite{bib-MUL00} in the Stranski-Krastanov
growth mode. In this context Henri \surname{Mariette} proposes an interesting contribution:
\emph{Key parameters for the formation of self-assembled quantum dot induced by the
Stranski-Krastanov transition: a comparison for various semiconductor
systems}\cite{bib-CRphys-Mariette}. A surprisingly simple model allows one to understand
quantitatively the occurrence of semiconductor quantum dots, by taking into account both the
strain decrease in the dots and conventional surface energy arguments occurring in the growth
modes recalled above. Such understanding is essential to increase the versatility of \SA. An
original aspect of the model is to explain the effect of parameters other than just material,
substrate, temperature and amount of material deposited, namely the use of surfactants to
manipulate the surface free energy and thus control the formation and density of quantum dots.

The advantage of \SO over \SA was mentioned in the introduction: the dispersion of size is greatly
reduced, thus either easying fundamental studies over assemblies of nearly mono-disperse objects,
or improving materials for applications by reducing the dispersion of physical properties~(the
order may also allow one to apply the powerful analytical methods of diffraction, see the
contribution by T. Metzger \etal just below). \SO\ generally does not arise during deposition on
crystalline surfaces, because once nucleated nanostructures are often immobile, as
substrate-deposit interactions are much smaller than interactions between neighboring
nanostructures~(this contrasts with ordered arrays of clusters fabricated by chemical means, see
below the contribution of B.~Chaudret). Thus in most cases the organization of deposited
nanostructures relies on the nucleation in an ordered fashion, in registry with a pre-existing
regular pattern on the growth surface. The way this pre-existing pattern gets organized should in
principle be the step that should first be called self-organized. Sylvie \surname{Rousset} \etal
propose an overview of the issues related to self-organization of surfaces including steps,
intrinsic and adsorbate-driven reconstructions, as well as the use of such surfaces for
self-organized overgrowth, mainly illustrated by results of their group\cite{bib-CRphys-Rousset}.

New fabrication processes often trigger the need for new characterization techniques. The most
widely reported studies of \SO systems are performed in real space, using electron or scanning
probe microscopies. A complementary approach is X-ray scattering. Indeed, thanks to coherence
lengths of typically a few micrometers, information on electron density fluctuations on a larger
length scale than atomic spacings can be obtained with a high precision using powerful analytical
tools. In principle, the entire mesoscopic range can be covered by x-ray scattering, from the
order of nanostructures down to their atomic structure. Besides, X-ray techniques are
non-destructive and can be performed \insitu. Mainly small-angle scattering\cite{bib-GUI95} and
satellite diffraction studies are reported in the literature, as reported in recent
reviews\cite{bib-SCH04,bib-SCH04b,bib-STA03}. An example of such studies is reported by Till
\surname{Metzger} \etal, entitled \emph{X-ray characterization of self-organized semiconductor
nanostructures}\cite{bib-CRphys-Metzger}. These authors present an original combination of grazing
incidence techniques, \ie, grazing incidence diffraction, with anomalous dispersion effects, \ie,
at energies close to an adsorbtion edge of one of the constituent elements, to add chemical
sensitivity to the technique. Applied to the case of SiGe quantum dots, this allows the chemical
composition and strain to be determined, a so-called \emph{iso-strain-scattering} technique. Such
studies are essential as physical properties like luminescence critically depends on composition
and strain.

Semi-conductor nanostructures play a central role in \SA and \SO. As mentioned above these
triggered the first growth investigations with the prospect of high-efficiency lasing in
nanometer-sized dots\cite{bib-ARA82,bib-ASA86}. Since that time the prospects have been refined.
Quantum wires or quantum dots will certainly not replace quantum wells because their uniformity in
size is not sufficient. Foreseen applications rather consist of niche applications, like
single-photon sources, or quantum-dot lasers with a wavelength in a range not achievable with
quantum wells. Several recent reviews are available on the
subject\cite{bib-REI03,bib-BER03b,bib-MIC03} and no contribution was included in the present
Issue.

Magnetism is another field that has benefited from \SO and \SA systems. The studies started later
than for semiconductors, with the first growth demonstrations published at the beginning of the
nineties. As the field is becoming riper, systems are now not solely grown for demonstration, but
their properties are investigated for specific purposes since a few years, with a clear rising
interest and number of reports. This interest is stimulated by the fact that the further increase
of density of magnetic recording media seems to come to an end using conventional film media.
Similarly, magnetic elements are now integrated in magneto-electronic devices like magnetic random
access memories~(MRAMs), for which the trend of miniaturization raises similar issues as for
recording media. Olivier \surname{Fruchart} proposes a review article entitled \emph{Epitaxial
self-organization: from surfaces to magnetic materials}\cite{bib-CRphys-Fruchart}. This consists
of an overview of the use for magnetic purposes that has been done up to now of \SO and \SA, with
an emphasis on aspects that could not have been addressed with nanostructures fabricated by
lithography. A first set of issues concerns the quantitative study of low-dimensional magnetic
phenomena~(1D and 0D) that also occur in functional materials but cannot be studied because of
their complexity: magnetic ordering, magnetic anisotropy, superparamagnetism. A second set of
issues concerns the possible direct use of self-organized systems in devices. Examples are given
how superparamagnetism can be fought, and what new or improved functionalities can be expected.

A magnetic achievement made possible by \SO and \SA is presented in more detail in the manuscript
by Pietro \surname{Gambardella} \etal, who report on \emph{magnetic anisotropy from single atoms
to large monodomain islands of Co/Pt(111)}\cite{bib-CRphys-Brune}. Magnetic anisotropy is an
essential property because it allows the freezing of magnetic systems in a given direction, and
therefore is a keystone of magnetic recording. Understand how this anisotropy is modified and can
be manipulated in nano-sized systems is a challenge, that can be tackled using \SO and \SA to
produce model systems for fundamental investigations. These authors studied model systems
consisting of large flat islands, stripes narrow down to mono-atomic wires, and controlled
clusters down to isolated atoms. Thus, the evolution of magnetic anisotropy from bulk crystals to
isolated atoms was fully covered, \ie, from 3D to 0D. The physical conclusion is that magnetic
anisotropy mainly arises from the atoms that posses the lowest dimensional coordination in the
system: surfaces for ultrathin films, which was already predicted\cite{bib-NEE54}, checked
experimentally\cite{bib-GRA68}, understood microscopically\cite{bib-BRU89b} with again
experimental confirmation\cite{bib-WEL95}. Here the conclusion is extended to edges for OD and 1D
objects. This new piece of knowledge opens the perspective to control the magnetic anisotropy
independently from the size of the system by engineering the area of interfaces in a compact
cluster, or of the length of edges in a flat structure, in an analogous approach to the control of
the dimension in fractal structures. This is illustrated by an original example of engineered
growth, were flat rings of Co were fabricated by step-decoration of non-magnetic Pt flat islands.

A prerequisite for the use of spontaneously-ordered systems in devices is to increase the
versatility of patterns that can be achieved, and also the quality of the order. Growth studies in
this direction are presented in the next two manuscripts. First G{\"u}nther \surname{Springholz}
proposes a literature review of the the \emph{three-dimensional stacking of self-assembled quantum
dots in multilayer structures}\cite{bib-CRphys-Springholz}. It was indeed shown that dots from
successive layers can display correlations, so that a three-dimensional ordered superstructure can
be fabricated upon deposition of many layers of dots and spacer
layer\cite{bib-XIE95,bib-SPR98,bib-TER96}, although the dots from the first layer are only
self-assembled. The manuscript reviews the various physical effects that determine the occurrence
of this order, the range of geometrical and material parameters required for the ordering, the
type of stackings that can result, and ways to control the super-lattice parameters. As stressed
in the introduction, ordering is interesting because it reduces the dispersion of size, thus of
physical properties. \SO three-dimensional stackings could be used in a very general way: if a
stacking is inert with respect to the physical properties that are sought, it could be used as a
template for the overgrowth of the material of physical interest, \ie, the stacking could be a
building block of a more complex stacking\cite{bib-CAP03}. New physical properties may also emerge
because of interactions between dots through the spacer layers.

Although stackings increase the versatility of \SO, still only a restricted number of patterns can
be obtained, that are all regular. In order to lift this limitation but still benefit from the
advantages of spontaneous fabrication under ultra-high vacuum, a new direction of research
consists in combining a first step of artificial structuring to impose a pattern, not necessarily
regular, with a second step consisting of growth self-organized in registry with the pre-defined
patterns. The main approaches followed are reviewed by their developers in the manuscript of Jo{\"e}l
\surname{Eymery} \etal, entitled \emph{nanometric artificial structuration of semiconductor
surfaces for crystalline growth}\cite{bib-CRphys-Eymery}. These consist of the use of arrays of
buried dislocations produced by controlled wafer bonding~(fast and with a low pitch, however not
more versatile than regular arrays), lithography patterning, or ion implantation patterning. These
studies initiated in the field of semiconductors about five years ago, and were eased by the
existing technological procedures to prepare again clean epitaxial surfaces after the
pre-patterning step. More recently a few reports appeared for the control of nucleation of
metals\cite{bib-YU01,bib-CHE04}, or clusters\cite{bib-BAR02,bib-GIE03}.

The Special Issue is concluded by a manuscript that does not relate epitaxial systems, but
nanoparticles synthesized by chemical means. Physical chemistry is indeed a promising route route
to yield both model systems for fundamental studies, and materials for applications. Many
approaches have been developed to yield nanoparticles by chemical means\cite{bib-LIZ04}, and let
them self-organize in 2D or 3D arrays. A first advantage of chemistry over physical deposition is
the lower magnitude of interactions between molecules than between epitaxial nanoparticles, which
favors the ordering. A second advantage is the nearly infinite possible combination of ligands,
that give many more degrees of freedom than the choice of elements for epitaxial growth. A popular
chemical route for the fabrication of magnetic nanoparticles is inverse
micella\cite{bib-PIL02,bib-LAL04,bib-TON98}. In this Issue Bruno \surname{Chaudret} presents an
original route recently opened, \ie, an \emph{organometallic approach to nanoparticles synthesis
and self-organization}\cite{bib-CRphys-Chaudret}. As detailed in the article, this approach
presents several advantages. First, it benefits from the existing know-how in metal-organic
chemistry, yielding metals, semi-conductors and oxides, each with a great variety of materials, in
pure clusters or core-shell structures. Second and probably the most original feature of this
report, the shape of the clusters can be controlled, \eg, to yield cubes. In SO 2D and 3D
supra-crystals of these clusters, this anisotropic shape can induce the alignement of the
crystallographic axes of all the clusters along a given direction, owing to nearest-neighbor
interactions via ligands. It is of prime importance when the physical properties of interest of
the particles are anisotropic, because in this case alignement of all major axes in given
directions can be required for integration into devices. This is for example the case for the
high-magnetic-anisotropy FePt nanoparticles that may be used in future generations of magnetic
recording media. The alignement of all axes reduces the distribution of magnetic switching field
mandatory to achieve sharp bit transitions thus reducing the number of grains per bits to its
minimum, as well as provide a nearly $100\%$ remanence, desirable for maximizing the read-out
signal and the SNR. The anisotropic shape easily obtained is explained by the anisotropic role of
the ligands during growth, while only the anisotropic surface energy of the crystalline clusters
plays a role in inverse micella. The organo-metallic approach is also complementary to inverse
micella because it allows the synthesis of smaller clusters, typically a few nanometers in
diameter against several tens of nanometers, respectively.

To conclude, self-assembly and self-organization on surfaces has become an established field in
the community of growth, with a steep rise of attention in the last five years. Whereas
self-organization was initially motivated by potential applications, up to now it has essentially
contributed to the advancement of fundamental science in low dimensions, for example in the fields
of semiconductors and magnetism, as model systems could be produced, that could not have been
fabricated by lithography. However there are still breakthroughs under way in extending the
versatility of self-organization, with a great hope in the emerging field of the combination of
first an artificial pre-patterning, followed by self-organized growth in registry with the
artificial pattern. This lifts the limitations of self-organization regarding the regularity of
the pattern, while keeping model nanostructures. Thus, self-organized systems are still under
active development and may anyhow find technological applications in the future.



\section*{Acknowledgements}
I would like to express my gratitude to Jacques Villain, who has given me the opportunity
to coordinate this Special Issue.

\bibliographystyle{report-nodescription}
\bibliography{fruche4,fruchart}

\begin{thebibliography}{10}
\bibitem{bib-CRphys-editor-SOSA}
\texttt{We will abusively use the acronym \SO~(resp. \SA) to stand for
  different expressions like self-organization, self-organize, self-organized.}
\end{thebibliography}







\end{document}